\documentclass[aps,prl,twocolumn,amsmath,amssymb,superscriptaddress,floatfix,showpacs,times]{revtex4-1} 
\pdfoutput=1
\usepackage{graphicx}
\usepackage{CJK}
\usepackage{times}

\usepackage{color}

\begin{document}

\begin{CJK*}{UTF8}{gbsn}

\title{Berry phase induced dimerization in one-dimensional quadrupolar systems}

\author{Shijie Hu (胡时杰)}
\affiliation{Max-Planck-Institut f\"ur Physik komplexer Systeme, 01187 Dresden, Germany}
\author{Ari M. Turner}
\affiliation{Department of Physics and Astronomy, The Johns Hopkins University, Baltimore, Maryland 21218}
\author{Karlo Penc}
\affiliation{Institute for Solid State Physics and Optics, Wigner Research Centre for Physics, Hungarian Academy of Sciences, H-1525 Budapest, P.O.B. 49, Hungary}
\author{Frank Pollmann}
\affiliation{Max-Planck-Institut f\"ur Physik komplexer Systeme, 01187 Dresden, Germany}

\date{\today}

\begin{abstract}
We investigate the effect of the Berry phase on quadrupoles that occur for example in the low-energy description of spin models.
Specifically we study here the one-dimensional bilinear-biquadratic spin-one model. 
An open question for many years about this model is whether it has a non-dimerized fluctuating nematic phase. The dimerization has recently been proposed to be related to Berry phases of the quantum fluctuations.
We use an effective low-energy description to calculate the scaling of the dimerization according to this theory, and then verify the predictions using large scale density-matrix renormalization group (DMRG) simulations, giving good evidence that the state is dimerized all the way up to its transition into the ferromagnetic phase.
We furthermore discuss the multiplet structure found in the entanglement spectrum of the ground state wave functions.
\end{abstract}


\maketitle

\end{CJK*}
% Introduction
How quantum fluctuations melt a classical order and create novel quantum states is a fundamental question of modern condensed matter physics.
Among many phenomena, mechanisms involving topological effects \cite{mermin} have been studied in great detail in one-dimensional spin chains as minimal models \cite{haldane4, affleck1}.
In particular, the Berry phase associated with rotation of spins plays a crucial role \cite{Berryphase}, as its presence discriminates between antiferromagnetic Heisenberg chains with half-integer and integer spins, making the excitations in the former gapless and in the latter gapped \cite{haldane1}.
\begin{figure}[t]
\includegraphics[width=\columnwidth]{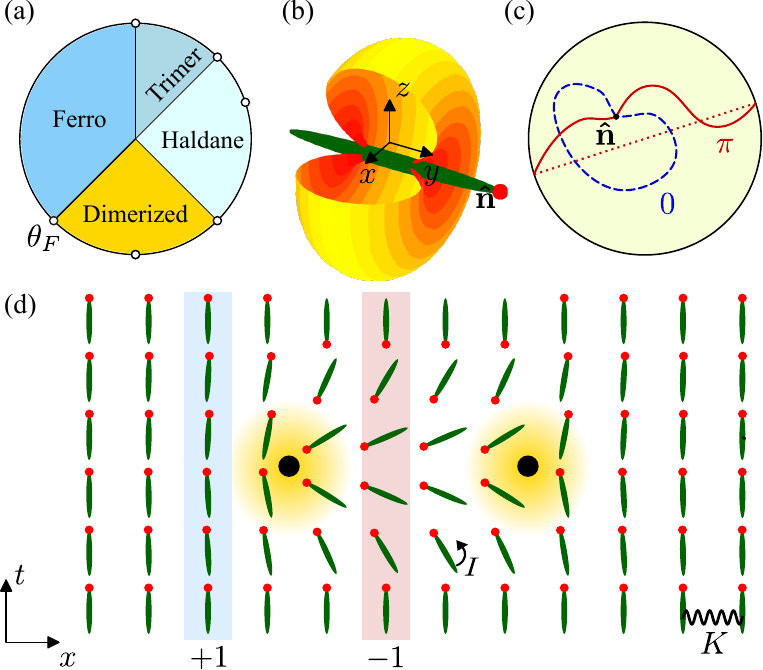}\\ 
\caption{(color online)
(a) Phase diagram of the 
Hamiltonian (\ref{ham}).
(b) In the quadrupolar state $|y\rangle$ the spin fluctuates over the yellow region perpendicular to the director (green) \cite{andreas2006,barnett2006}.
(c) Rotating a director in the projective plane, a Berry phase is picked up on the red path only (the dashed red line connects equivalent points in $\mathbb{RP}^2$).  
(d) World lines of directors in $1+1$ dimensional space time.
The directors tend to align with adjacent ones in space (due to the spring constant $K$), and they can rotate in time, with a moment of inertia $I$.
Rotations of a director as a function of time lead to a Berry-phase. Directors can either rotate by an angle $0$ or $\pi$ as shown in the two highlighted world lines, picking up a phase $\pi$ in the latter case.
To keep track of the total rotation, we add a red dot onto the green ellipsoid.  
In between domains of different windings, $\pi$-vortices appear (black dots).
}
\label{fig1}
\end{figure}

A similar Berry phase appears for the spin quadrupoles (nonmagnetic spin states)  we investigate here.  
The quadrupolar order can be found in the $S=1$ Heisenberg model
\begin{equation}
{\cal H}_{\text{BB}} = \sum_{j=1}^{L-1} 
\cos\theta \left({\mathbf S}_{j}\cdot{\mathbf S}_{j+1}\right)
+ \sin\theta \left({\mathbf S}_{j}\cdot{\mathbf S}_{j+1}\right)^2 \;,
\label{ham}
\end{equation}
where $\theta$ parameterizes the ratio of the bilinear and biquadratic terms and $L$ is the length of the chain.
The phase diagram of this model is shown in Fig.~\ref{fig1}(a) \cite{chubukov3,*chubukov2}. 
It is generally agreed that the model exhibits a ferromagnetic phase, a gapped ``Haldane'' phase \cite{haldane1}, a gapless trimerized phase and a gapped dimerized phase \cite{barber,*klumper,*xian}. 
A long lasting debate has been going on about the possible existence of a non-dimerized gapped phase above the $\theta_F = -3\pi/4$ SU(3) symmetric point.
This phase can be thought of as a nematic phase that has become disordered on account of quantum fluctuations (since the system is one dimensional).
At $\theta_F$ the ground state is degenerate, and it has no fluctuations: any state where each spin is in the same state is a ground state.
Slightly away from $\theta_F$, nematic states of the spins have lower energies than other states, so the state will be approximately a uniform state of nematic spins.
Due to quantum fluctuations, the order will only last up to the correlation length.
There is no obvious reason that this state should become modulated with a period of $2$ near this point (there is no minimum in the spin-wave dispersion near $\pi$).
Could this be a fifth homogeneous phase? % between the ferromagnetic and the dimerized phase?
This question was first raised in \cite{chubukov3,*chubukov2}, which also calculated that this phase would exist for $\theta_{F} < \theta \lesssim-0.66\pi$ \cite{chubukov3,*chubukov2}.
This idea has attracted considerable interest \cite{ivanov,kawashima, laeuchli,solyom3,yip}, and despite the progress in numerical techniques, more recent simulations still are producing contradicting results \cite{rizzi,*solyom1, *orus, *porras}.

Refs. \cite {moessner, grover} found a reason why the nematic phase becomes dimerized, related to Berry phases associated with quantum fluctuations of the quadrupoles, which we will test here.
We will find how the dimerization near $\theta_F$ varies using a quantum rotor model, and then calculate the actual dimerization numerically to check the prediction.
We further explore the entanglement spectrum for which we derive exact expressions at $\theta_F$.

Let us start from the ordered quadrupolar mean-field wave function in the vicinity of $\theta_{F}$ with  $\theta \ge \theta_{F}$:
\begin{equation}
\left|\psi_{\text{MF}}\right\rangle\!=\!\prod^{L}_{j=1} \left|\psi\right\rangle_{j}\!=\!\prod^{L}_{j=1}\left(
 n^{x} \left| x \right\rangle_j\!+\!n^{y} \left| y \right\rangle_j\!+\!n^{z} \left| z \right\rangle_j \right),
\label{MF}
\end{equation}
where the coefficients are the components of the ${\hat {\mathbf n}} = (n^{x},\ n^{y},\ n^{z})$ real unit vector.
The states $|x\rangle_{j}\!=\!{\it i} (|1\rangle_{j}\!-\!|\bar 1\rangle_{j})/\sqrt{2}$, $|y\rangle_{j}\!=\!(|1\rangle_{j}\!+\!|\bar 1\rangle_{j})/\sqrt{2}$ and $|z\rangle_{j}\!=\!-{\it i} | 0\rangle_{j}$ form the time-reversal invariant basis of the $S=1$ spins at site $l$ ($|1\rangle$, $|0\rangle$ and $|{\bar 1}\rangle$ are the $S^{z}$ eigenstates).
Being fully symmetric, $\left|\psi_{\text{MF}}\right\rangle$ is an exact ground state at the SU(3) symmetric point $\theta_{F}$.
The $\left|\psi_{\text{MF}}\right\rangle$ does not have the full SO(3) symmetry. The order parameter can be thought of as an ellipsoid, as shown in Fig.~\ref{fig1}(b).
It has a rotational symmetry 
$\exp({\it i}\gamma{\hat {\mathbf n}}\cdot{\mathbf S})|\psi_{\text{MF}}\rangle=|\psi_{\text{MF}}\rangle$ 
with any angle $\gamma$ around 
${\hat {\mathbf n}}$, as well as an additional ${\mathbb Z}_{2}$ symmetry from flipping the direction axes ${\mathbf{\hat{n}}} \rightarrow -{\mathbf{\hat{n}}}$.
This extra symmetry distinguishes the quadrupolar order from the ferromagnetic order. As a consequence, the ``director'' $\mathbf{\hat{n}}$ lives in the projective plane $\mathbb{RP}^2$ formed by identifying antipodal points of the unit sphere. 
Because of the $\mathbb{Z}_2$ symmetry, we can distinguish two topologically distinct classes of closed adiabatic paths of the $\mathbf{\hat{n}}$: the paths can cross the boundary of the $\mathbb{RP}^2$ an even or odd number of times [see Fig.~\ref{fig1}(c)].
Since the Berry phase for the time reversal invariant quadrupoles is quantized to $0$ or $\pi$  \cite{Hatsugai06}, the phase on homotopic paths is equal.
In the case of 
%$0$ or 
even number of crossings the path can be contracted to a single point, and we expect no Berry phase. For an odd number of crossings the path cannot be contracted, and the wave function 
acquires a Berry phase $\pi$ for a spin-$1$ quadrupolar state.
To see this, we rotate the $|z\rangle$ around the $y$-axis by changing $\varphi = 0$ to $\pi$ as $|\psi(\varphi)\rangle=\cos \varphi |z\rangle + \sin\varphi |x\rangle$. 
In this gauge the Berry connection $i \langle\psi(\varphi)|\partial_{\varphi} |\psi(\varphi)\rangle =0$ and the Berry phase is just given by the change of the sign of  the wavefunction which is $-1$.

On account of the Mermin-Wagner theorem, quantum fluctuations are large enough to cause the state to become disordered.
To understand the consequences, we will assume that the low energy fluctuations are described by a rotor model\cite {ivanov,lamacraft}
\begin{eqnarray}
{\cal H}_\text{rotor}=\sum_{j=1}^{L}\frac{{\mathbf L}^{2}_{j}}{2I}-\frac{K}{2}\left({\hat {\mathbf n}}_{j}\cdot{\hat {\mathbf n}}_{j+1}\right)^{\!2}.
\label{HAM}
\end{eqnarray}
Here the first term is the kinetic energy of the directors (the original spin ${\mathbf S}_{l}$ and the angular momentum ${\mathbf L}_{j}=I\ {\hat {\mathbf n}}_{j}\times{\partial_{t}{\hat {\mathbf n}}_{j}}$ coincide when averaged over enough sites), the second term is the $\mathbb Z_{2}$ symmetric interaction between two adjacent directors.
The energy of a twist in the director and the response to a magnetic field can be calculated both from the rotor model and the original spin model (using the mean-field approximation) \footnote{See Supplemental Material for the derivation of $K$ and $I$.}, and comparing these tell us the spring constant and moment of inertia, $K\!=\!-2\sin\theta$, and $I\!=\!1/2(\cos\theta\!-\!\sin\theta)$ \cite{ivanov,lamacraft}.
In the continuum description of the model, we rescale the imaginary time $t={\tilde t}\sqrt{I/K}$ so that the action becomes isotropic
\begin{equation}
\mathcal{S} = \frac{1}{2g_0} \iint d{\tilde t} d{\mathbf x} \left[\left(\partial_{\tilde t} \hat {\mathbf n}({\tilde t},{\mathbf x})\right)^{\!2}\!+\!\left(\partial_{\mathbf x}\hat {\mathbf n}({\tilde t}, {\mathbf x})\right)^{\!2}\right],
\label{eq:contact}
\end{equation}
where the dimensionless stiffness $1/g_0=\sqrt{IK}$.

We now use space-time path integrals to show that, when the Berry phase associated to the $\mathbb{Z}_2$ symmetry of Eq.~(\ref{eq:contact}) is $\pi$ (as for the spin-$1$ case) the ground state becomes dimerized.  
As illustrated in Fig.~\ref{fig1}(d), the $\mathbf{\hat{n}}$ field in the two-dimensional $(x,\tilde {t})$ space has topological defects in the form of \emph{vortices}, characterized by the fundamental group $\pi_1(\mathbb{RP}^2) = \mathbb{Z}_2$ \cite{nematic_defects_review_2012}.
Because of the continuous O(3) rotation symmetry $\mathbf{\hat{n}}$ does not have long-range order.
Since $\mathbf{\hat{n}}$ has three components, just spin waves lead to exponentially decaying correlations \cite{polyakov} (vortices are not necessary).
This causes vortices to have a finite action, so ``entropy'' produces them at any stiffness, making the entire phase dimerized.
That is because the vortices separate domains where worldlines of the directors have different winding numbers (and hence different Berry phases $0$ and $\pi$).
Let us for clarity first assume that we have only one pair of vortices.
The sign of the path integral depends on their distance: the sign is positive (negative) if the vortices are separated by an even (odd) number of sites.
For $2N^{v}$ vortices, the product over all pairs gives the total sign $\prod^{2N^{v}}_{j=1} (-1)^{x^{v}_{j}}$, where $x^{v}_{j}$ are the indices of bonds that vortices live on.
This gives rise to a dimerization of the system as path integrals come with opposite signs depending on whether the vortices are located on even or odd bonds relative to each other.
From this reasoning, we expect that the dimerization strength ${\mathcal D}$ is proportional to the density of vortices.
We now apply this qualitative theory and predict the scaling of the correlation length and dimerization strength.
Based on the fact that the O(3) symmetry  is not broken, we find, as noted in Ref.~[\onlinecite{polyakov}], that the renormalized stiffness $1/g(r)\!=\! 1/g_0\!-\!(\ln r)/2\pi$ decreases for longer length scales (the distance $r^{2}\!=\!x^{2}\!+\!{\tilde t}^{2}$), eventually disappearing for $r \agt \xi$, where
\begin{equation}
\xi\!=\!\exp\left(2\pi\sqrt{IK}\right)\!\approx\!\exp\left(\pi \sqrt{2/\Delta\theta}\right)
\label{eq:correlationlength}
\end{equation}
and $\Delta\theta\!=\!\theta\!-\!\theta_{F}$.
The correlations also disappear for $r\agt\xi$. 
This derivation is very similar to the argument in Ref.~[\onlinecite{ivanov}], up to a factor of two in the exponential. 

Supposing that the dimerization $\mathcal{D}$ is proportional to the density of vortices, we are now able to estimate the dimerization strength quantitatively.
For a classical vortex shown in Fig.~\ref{fig1}(d), we find that $|\nabla {\hat {\mathbf n}}|^{2} = 1/4r^{2}$. To take into account that the size of a vortex is $\xi$, we assume that the stiffness is decreasing gradually from the core of the vortex out to $\xi$, so that the action $\mathcal{S}^{v}\!=\!\int^{\xi}_{1} d{r} 2\pi r  |\nabla {\hat {\mathbf n}}|^{2}/2g(r)\!=\!\left(\ln\xi\right)^2\!/16$.
The density of vortices is proportional to $\exp(-{\mathcal S}_{v})$ and thus
\begin{equation}
\mathcal{D}\!\propto\!\exp\left(-\frac{\ln^2 \xi}{16}\right)\!\approx\!\exp\left(-\frac{\pi^2}{8\Delta \theta}\right) \;.
\label{eq:dimer}
\end{equation}
The gap has the same order as $1/\xi\approx \exp(-\pi\sqrt{2/\Delta\theta})$, so the dimerization is much smaller than the gap close to $\theta_F$.
This is very different than usual behavior seen in spin chains: for example, in the frustrated spin--1/2 Heisenberg chain, the spontaneous dimerization is $\mathcal{D}\propto1/\sqrt{\xi}\propto \sqrt{\Delta_\text{dim}}$ \cite{white96}.

\begin{figure}[t]
\includegraphics[width=\columnwidth]{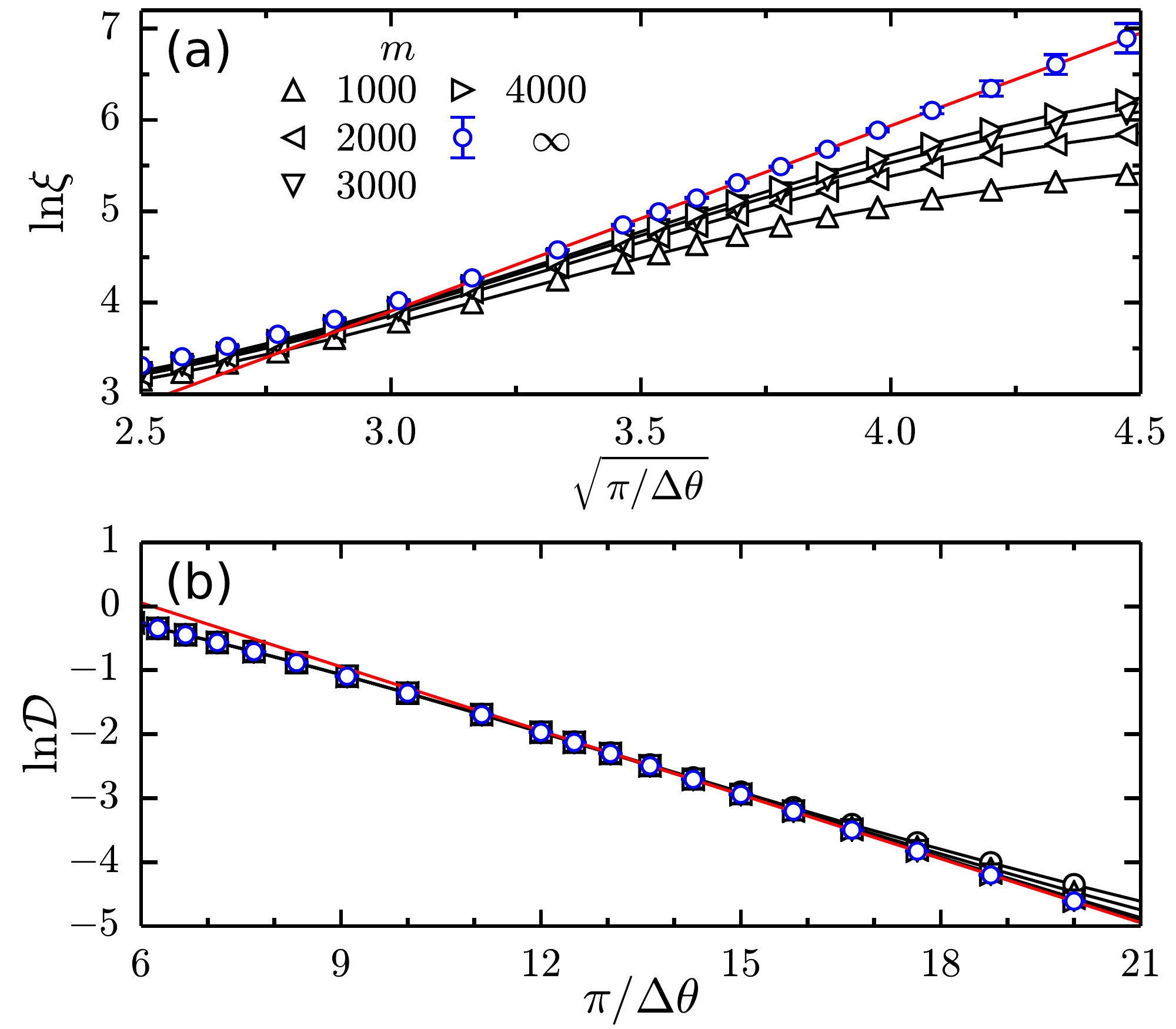}
\caption{(color online)
(a) the correlation length $\ln\xi$ and (b) the bulk dimerization $\ln{\cal D}$ as a function of $1/\sqrt{\Delta \theta}$ and $1 /\Delta \theta$ respectively.
The data is extrapolated to the infinite-$m$ limit (blue open circles with error bars).
Red fitting lines: $\ln\xi\!=\!-2.180 + 3.596 /\sqrt{\Delta \theta}$ in (a) and $\ln{\cal D}\!=\!2.054 - 1.047/\Delta \theta$ in (b).
}
\label{fig2}
\end{figure}

We will now compare expressions (\ref{eq:correlationlength}) and (\ref{eq:dimer}) with numerical DMRG simulations \cite{white1,white2}.
In order to reach sufficiently large system sizes, we use an improved DMRG technique with open boundary conditions \cite{shijie}:
We performed simulations with up to m = 4000 kept states for system sizes up to $L=20000$ and extrapolate to the $m\rightarrow\infty$ limit where needed.
The correlation length is obtained by diagonalizing  the transfer matrix \cite{mcculloch} and the dimerization ${\mathcal D}$ measures the energy difference between strong and weak bonds (the open chain with even sites dimerizes).
Our results, shown in Fig.~\ref{fig2}, confirm the scaling behavior of both the correlation length and the dimerization predicted by the effective field theory in the vicinity of $\theta_F$. 
Based on numerical simulation we conclude that the low-energy (i.e., long wavelength) behavior is described by the rotor model together with Berry phases, yielding a gapped and dimerized phase up to the transition to the ferromagnetic phase.
After focussing on the physics at long distances, we will now also consider shorter length scales. 
\begin{figure}[t]
\includegraphics[width=\columnwidth]{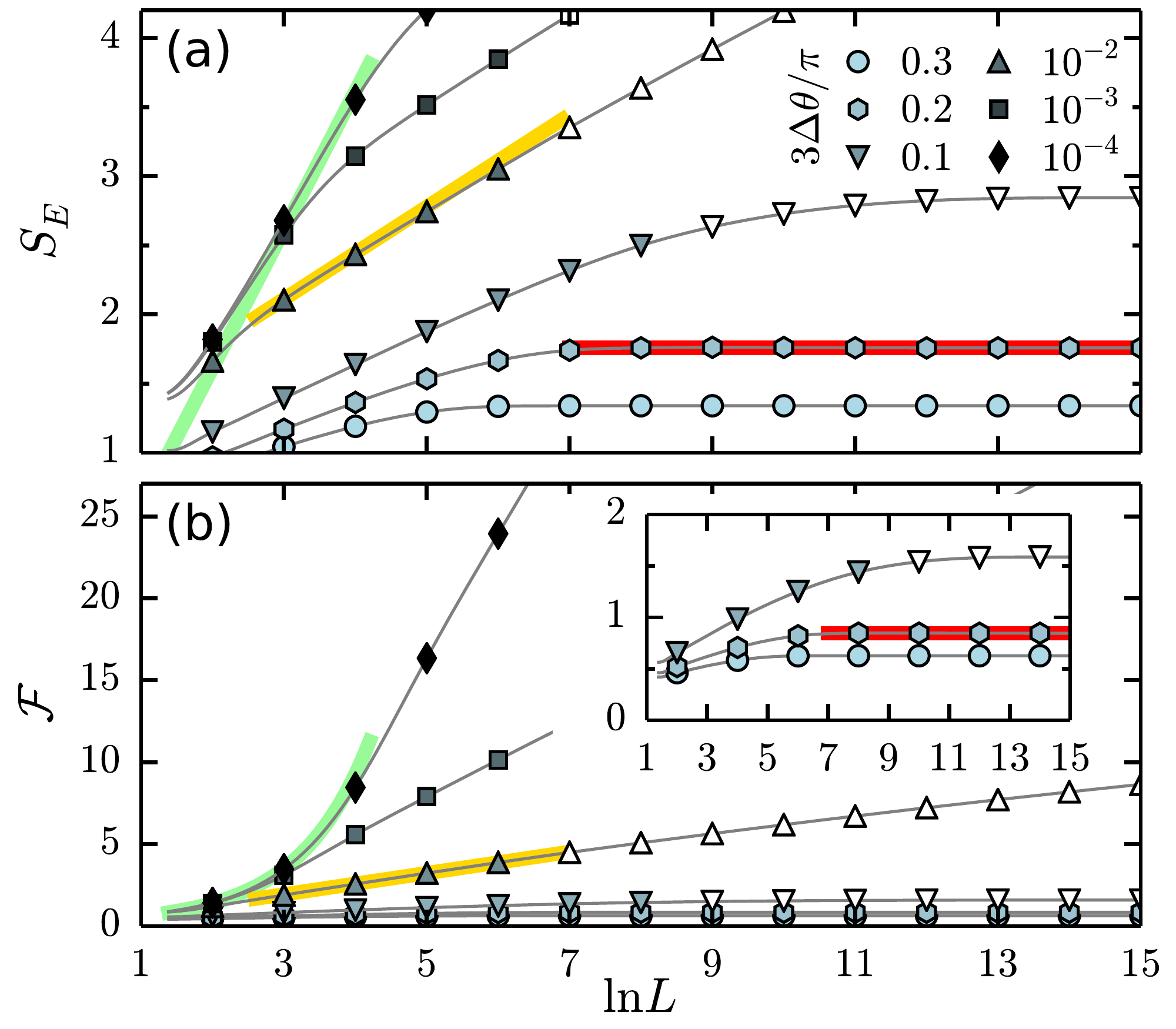}
\caption{(color online)
Three different regions can be distinguished by the scaling of the (a) entanglement entropy $S_{E}$ and (b) bipartite spin-fluctuation ${\cal F}$ as a function of $\ln L$ for different values of $\Delta \theta/\pi$.
Converged (interpolated) data are marked by filled (open) symbols.
Fat green, yellow and red lines indicate the expected asymptotic behavior in the three regimes (see text for details).
}
\label{fig3}
\end{figure}
We find the three different regions shown in Fig.~\ref{fig3}. 
The bipartite spin fluctuations are given by ${\mathcal F}\!=\!\langle \left(S^{z}_{\mathcal L}\right)^{\!2}\rangle$ \cite{song} and the $S_E$ is the von Neumann entropy of the reduced density matrix $\rho_{\mathcal{L}}$. 
Both quantities are calculated for the bipartition of a chain of length $L$ into two (left, $\mathcal{L}$, and  right, $\mathcal{R}$) half chains of equal length. 
At short and intermediate distances, the ground state appears to be critical. 
In particular, the entanglement entropy $S_E$ shows a logarithmic growth as a function of system size.
This is a known hallmark of conformally invariant, critical systems for which $S_E = (c/6) \log L$ with $c$ being the central charge, i.e., the number of linearly dispersing modes \cite{blote,calabrese}.
Though our system is \emph{not} conformally invariant, we still use the notation of $c$ to characterize the behavior at shorter length scales.
The short distance region is described by an effective $c=6$ and the bipartite fluctuations $\mathcal{F}$ grow linearly with the system size (green lines).  
Both properties are inherited from the exactly soluble SU(3) point. 
Above a scale $L_m\propto 1/\sqrt{\Delta\theta}$, the \emph{magnetic healing length}, but below $\xi$, the critical behavior changes to an effective $c=2$ and a logarithmic growth of $\mathcal{F}$ (yellow lines), because there are two approximate Goldstone modes of the rotors
\footnote{
Close to the $SU(3)$ point, the spin-wave dispersion $\propto k \sqrt{8 \Delta \theta + k^2}$ shows a crossover behavior at $k_m \approx \sqrt{8\Delta\theta}$: for $k\ll k_m$ it is linear in $k$, while for $k \gg k_m$ the dispersion is quadratic \cite{chubukov3,*chubukov2,papanicolaou,karlobook}.
}.
%\footnote{This can be understood from the spin-wave analysis.
%%
%Close to the $SU(3)$ point, both the spin-wave dispersion $\propto k \sqrt{8 \Delta \theta + k^2}$ and spin correlations $\propto k/\sqrt{8 \Delta \theta + k^2}$ 
%%
%show a crossover behavior at $k_m \approx \sqrt{8\Delta\theta}$: for $k\ll k_m$ they are linear in $k$, leading to $c=2$ in the $S_E$ and $x^{-2}$ decay of spin correlations for $x\agt L_m \approx k_m^{-1}$,
%%
%while for $k \gg k_m$ the dispersion is quadratic with constant spin correlations \cite{chubukov3,*chubukov2,papanicolaou,karlobook}.
%}.
%
For $L\!\gg\!\xi$ the growth of the entropy is cut off by the gap and the fluctuations saturate (red lines).
The fact that the correlation length $\xi$ is of the order of thousands of sites in the scaling regime near $\theta_F$ might explain why previous numerical studies predicted a critical phase instead of a dimerized one. 

\begin{figure}[t]
\includegraphics[width=\columnwidth]{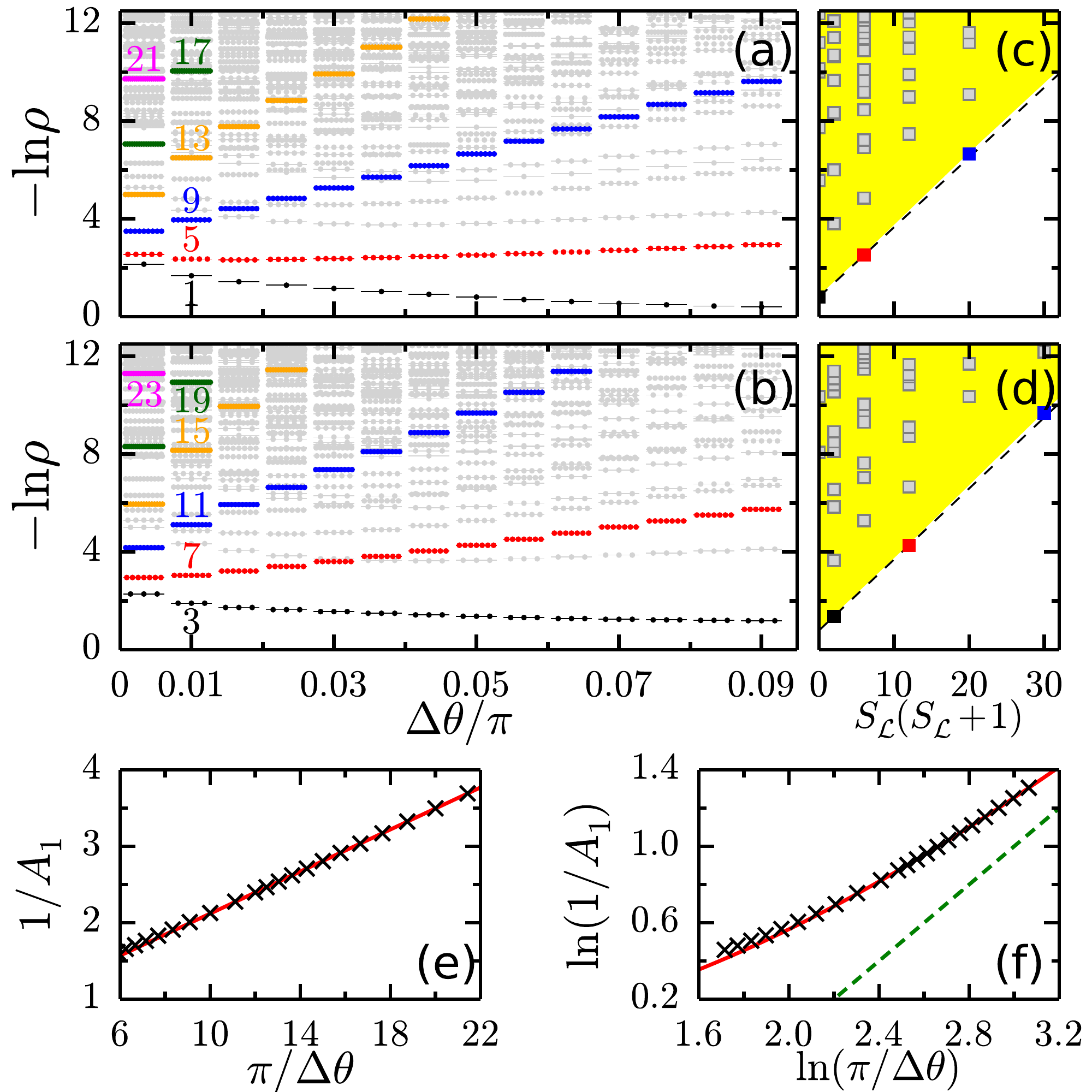}
\caption{(color online) Multiplet structures of the ES as a function of $\Delta \theta$ for  (a) even and (b) odd $L/2$.
Levels originating from the SU(3) point are colored and their multiplicity indicated.
(c) and (d) ES as a function of $S_{\mathcal L}(S_{\mathcal L}+1)$ for $\Delta\theta/\pi=0.05$. 
Black dashed lines $A_{0}+A_{1}S_{\mathcal L}(S_{\mathcal L}+1)$ are fitting to the lower edge of the sepctrum.
(a)-(d) are obtained with $m=4000$.
(e) The slope of tower edges as a function of $1/\Delta\theta$.
Each slope value is extrapolated to the infinite-$m$ limit with error bar less than the symbol size.
The red solid line gives $1/A_{1}=0.745+0.432/\Delta\theta$. 
(f) Double logarithmic plot with green dashed line showing the expected asymptotic slope of $1/A_{1} \propto 1/\Delta\theta$.  
}
\label{fig4}
\end{figure}
For a number of different systems it was shown that conclusive information about a state can be extracted from the \emph{entanglement spectrum} (ES) \cite{li,frank2,alba}, the set of eigenvalues $\{\rho_{\gamma}\}$ of the reduced density matrix $\rho_{\mathcal{L}}$.
Figures~\ref{fig4}(a) and (b) show the spectrum as a function of $\Delta \theta$ for the two inequivalent bonds.
The apparent multiplet structure reflects the SO(3) symmetry of the state.
The lowest levels have the same multiplicites as in a perfectly dimerized state, where cutting a strong bond gives a three fold degenerate state as we cut a spin-$1$ singlet, while cutting a weak bond gives a non-degenerate state.
In the limit  $\theta \rightarrow \theta_{F}^{+}$ we can actually derive an exact expression for the ES, as the ground state is the SO(3) singlet of the $(L,0)$ Young tableau.
The $2 S_{\mathcal L}+1$ degenerate eigenvalues of the reduced density matrix $\rho_{\mathcal L}$ for a bipartition into two half-chains are \footnote{See Supplemental Material for the derivation.}
\begin{eqnarray}
\label{eq:RSSSspectrum}
\rho_{S_{\mathcal L}}\left(L, \theta_{F}\right)\!=\left[\!\frac{2^{S_{\mathcal L}}\!\sqrt{L\!+\!1}(L/2)![(L/2\!+\!S_{\mathcal L})/2]!}{(L/2\!+\!S_{\mathcal L}\!+\!1)![(L/2\!-\!S_{\mathcal L})/2]!}\right]^2,
\end{eqnarray}
where the allowed spins $S_{\mathcal L}$ are even for  $L/2$ even and odd for $L/2$ odd. 
%$S_{\mathcal L}$ [Figs.~\ref{fig4}(a) and (b)].
%
The short-range behavior in Fig.~\ref{fig3}, $S_E\approx\ln L$ and $\mathcal{F}/L \approx L/6$,  is reproduced using Eq.~(\ref{eq:RSSSspectrum}).
%we get 
%$S_E/\ln L \rightarrow 1$  and $\mathcal{F}/L \rightarrow 1/6$ as $L \rightarrow \infty$.
%

%
As we move away from the $\theta\!=\!\theta_{F}$, additional states appear in the ES (grey levels in Fig.~\ref{fig4}).
Since the ground state is SO(3) symmetric, we can plot the ES as a function of $S_{\mathcal L}(S_{\mathcal L}+1)$, Figs.~\ref{fig4}(c) and (d).
The lower edge of the spectrum graphed in this way is linear, similar to the \emph{Anderson tower} found for the energy spectra of ordered spin systems \cite{Bernu2} (see also \cite{metlitski}). The presence of every other spin only is typical for a ferroquadrupolar phase \cite{papanicolaou,karlobook}. 
%(Anderson towers in entanglement spectra have also been discussed in {\cite{metlitski}}).
%
Unlike for gapless systems, the linear slope $A_1$ of the Anderson tower remains finite.
Its states contribute roughly $1/A_1$ to the spin fluctuation of an infinite half-chain ${\mathcal F}$
\footnote{The bipartite spin fluctuations are given by
\[
{\mathcal F} = \frac{
\sum_{S_{\mathcal L}} S_{\mathcal L} (S_{\mathcal L}\!+\!1) (2 S_{\mathcal L}\!+\!1)  \exp\left(-A_1 S_{\mathcal L} (S_{\mathcal L}\!+\!1) \right) }
{3 \sum_{S_{\mathcal L}} (2 S_{\mathcal L}\!+\!1) \exp\left(-A_1 S_{\mathcal L} (S_{\mathcal L}\!+\!1) \right)}
\!\approx\!
\frac{1}{3A_1}
\]
as $\theta \rightarrow \theta_{F}^{+}$.}. 
Thus we can estimate $A_1$ from the bipartite spin fluctuations if we assume other states do not change the order of magnitude of the spin fluctuations.
The spin correlation function is given by $G^{s}(x)=-(1/2\pi^2)\ln(\xi/x)/x^2$, so the bipartite spin fluctuations are ${\mathcal F}=-\int^{\xi}_{1} dx \int^{x-\xi}_{0} dy G^{s}(x-y)\propto \ln^2\xi$
\footnote{Since the ground state is a singlet, 
\mbox{$\langle\left( S^{z}_{\mathcal L}\right)^2\rangle=-\langle S^{z}_{\mathcal L} S^{z}_{\mathcal R}\rangle$}, which can be expanded as a sum of $G^{s}(x-y)$. To get a rough estimate, one can stop the sums when \mbox{$|x-y|> \xi$}.}
in the leading order.
Thus we expect $1/A_{1} \propto \ln^2\xi \propto1/\Delta\theta$, and this agrees with the numerical results shown in Fig.~\ref{fig4}(e) when $L\gg\xi$,
though the anticipated behavior is reached only in the asymptotic limit as indicated in Fig.~\ref{fig4}(f).

In this manuscript, we studied the effect of the Berry phase on quadrupoles that occur in the effective description of the one-dimensional bilinear-biquadratic spin-one model. 
From the effective low-energy theory, we showed that quantum fluctuations melt the nematic order to produce a gapped state, while
vortices and a Berry phase cause dimerization.
The scaling predicted by this theory was verified using large scale density-matrix renormalization group simulations.
Additional insight into the nature of the ground state was obtained by studying  the ES. 
Beyond magnetic systems, our findings are of relevance to cold spin one bosons, where dimerization would occur quite naturally \cite{yip}.

% Thanks %
\begin{acknowledgments}
We thank Xiaoqun Wang, Joe Bhaseen, Hong-Hao Tu and Jize Zhao for helpful discussions. This work was supported by Hungarian OTKA Grant No. 106047. K. P. gratefully acknowledges the hospitality of the guest program of MPI-PKS Dresden, where much of this work was carried out.
\end{acknowledgments}

\bibliographystyle{apsrev4-1}
%

%%%%%%%%%%%%%%%%%%%%%%%%%%%%%%%%%%
% Supplement 
%%%%%%%%%%%%%%%%%%%%%%%%%%%%%%%%%%%

\onecolumngrid
\bigskip
  \begin{center}
    \rule[1ex]{.5\textwidth}{1.0pt}
  \end{center}
\section{Supplementary Material}
%:\\Berry phase induced dimerization in one-dimensional quadrupolar systems}
\twocolumngrid

\renewcommand{\theequation}{S.\arabic{equation}}
\renewcommand{\thefigure}{S.\arabic{figure}}
\renewcommand{\thetable}{S.\Roman{table}}
\renewcommand{\thesubsection}{S.\arabic{subsection}}

\setcounter{equation}{0}
\setcounter{figure}{0}
\setcounter{table}{0}

\subsection{The coefficients of the rotor model}

In this section we describe how to calculate the spring constant and momentum of inertia in the rotor model defined by the ${\cal H}_\text{rotor}$ Hamiltonian, Eq.~(3) in the main paper.  The rotor model describes the energy of the fluctuations starting from the mean field solution of the spin-1 Heisenberg model, so we will first give some details on the mean field solution, and after that calculate the energy of the spatial and temporal fluctuations.
 
\subsubsection{Mean field wave function}

To describe the ferroquadrupolar state, we assume a site factorized wave function of the form
\begin{equation}
  |\Psi\rangle = \prod_{j=1}^L 
  \left(\nu^x_j |x\rangle_j + \nu^y_j |y\rangle_j + \nu^z_j |z\rangle_j \right) \;,
\end{equation}
where we allow for the coefficients $\bm{\nu}_j  = (\nu^x_j,\nu^y_j,\nu^z_j)$ to be complex numbers, furthermore we normalize the wave function, requiring $\langle \bm{\nu}_j | \bm{\nu}_j \rangle = \sum_{\alpha=x,y,z} \bar \nu^\alpha_j \nu^\alpha_j = 1$. The expectation value with the ${\cal H}_{\text{BB}}$ bilinear-biquadratic Heisenberg model [Eq.~(1) in the letter] is then
\begin{align}
\langle\Psi|{\cal H}_{\text{BB}}|\Psi\rangle  = \sum_{j=1}^{L} 
& \left[  
\left(\sin \theta- \cos \theta\right) \left|\langle \bm{\nu}_{j}|  \bm{\bar\nu}_{j+1} \rangle \right|^2 
   \right.
  \nonumber\\
 &
\left.
+ \cos \theta \left |\langle \bm{\nu}_{j}| \bm{\nu}_{j+1} \rangle \right|^2 
+ \sin \theta 
\right]
    \;.
   \label{eq:SHbbMF}
\end{align}
The mean field ground state when $-3\pi/4 < \theta < -\pi/2 $ is the ferroquadrupolar one,
where the $\bm{\nu}_j$ becomes the real ${\bf n}_j = {\bf n}$, with energy per site $\varepsilon_{\text{FQ}}= 2 \sin\theta$.

\subsubsection{Spring constant}
The coefficients of the effective Hamiltonian can be determined  by considering two types of excitation. In the space dimension the response to stresses at the ends of the chain is a wave where the directors rotate in a plane, e.g. in the $xy$ plane as $\bm{\nu}_j = (n^{x}_{j},n^{y}_{j},n^{z}_{j})= (\cos k j, \sin k j,0)$ with a characteristic wave number $k$. The 
$\left |\langle \bm{\nu}_j| \bm{\nu}_{j+1} \rangle \right|^2 =
  \left|\langle \bm{\nu}_j|  \bm{\bar\nu}_{j+1} \rangle \right|^2 = \cos^2 k$,
and from Eq.~(\ref{eq:SHbbMF}) we get that the energy cost per site of such an excitation is equal to 
\begin{equation}
 \varepsilon_{k}-\varepsilon_{\text{FQ}}= - (1-\cos^2 k) \sin \theta \;.
\end{equation} 
Note that $\varepsilon_{k}-\varepsilon_{\text{FQ}}$ is positive, as $\sin \theta<0$ for $\theta$ values of interest.

The energy per site of this excitation in the rotor model [Eq.~(3) in the letter]  is  $ \frac{K}{2} \left(1-\cos^{2} k\right)$, as ${\hat {\bf n}}_{j}\cdot{\hat {\bf n}}_{j+1} = \cos k$. Equating the two energies, we find that they agree if we choose 
\begin{equation}
 K = -2\sin\theta
\end{equation} 
 for the spring constant.

\subsubsection{Moment of inertia}

The other term is like kinetic energy, so consider a rotating state $|\Psi_\omega(t)\rangle$ that is rotating at a frequency, $\omega$. 
The energy changes because the rotation changes the state (similar to how in a rotating system, there are changes of the density from the centrifugal force). The change can be found
by using a rotating coordinate system;
 i.e. define $|\tilde{\Psi}_\omega(t)\rangle=e^{i\omega t \sum_j S^z_j}|\Psi_\omega(t)\rangle$.
Then $i\partial_t |\tilde{\Psi}_\omega(t)\rangle=(\mathcal{H}_{\text{BB}}-\omega\sum_j S^z_j)|\tilde{\Psi}_\omega\rangle$.
If the state is just rotating, then it is a stationary state, so it is an eigenstate of
\begin{equation}
 \mathcal{\tilde H} = \mathcal{H}_{\text{BB}} - \omega \sum_j S^z_j \;.
\end{equation}
The $\omega$ is like a magnetic field along the $z$ axis.% In fact, if we add a small external magnetic field to a system of quadrupoles, the %directors start to rotate with $\omega=h$ around the field. 

What is the energy? 
The pure quadrupolar state has vanishing matrix elements with the spin operators, thus it does not couple to magnetic field. However, once we apply the field, the 
$|\bm{\nu}\rangle$ deforms and the spins develop magnetic moments. The directors select a plane that is perpendicular to the field -- in this case, it is the $xy$ plane. We choose the wave function in finite field as $\bm{\nu}_j = (u^x,i v^y,0)$, so that $(u^x)^2+(v^y)^2=1$ for the real numbers $u^x$ and $v^y$. The energy per site in the presence of the Zeeman term (i.e. rotating frame) is modified to
\begin{equation}
 \varepsilon_\omega= 2 \sin\theta + 4 (\cos\theta-\sin\theta) (u^x v^y)^2 - 2 \omega u^x v^y \;.
\end{equation}
 Noting that $m=2 u^x v^y$ is the magnetization of the spin,  the energy becomes $\varepsilon_\omega= 2 \sin\theta +  (\cos\theta-\sin\theta)  m^2 -  \omega m $, with a minimum achieved for $m=\omega/2(\cos\theta-\sin\theta)$. The
 energy cost per site (we do not count the $-  \omega m$ term)  is then given as 
\begin{equation}
 \delta \varepsilon_\omega =   \frac{\omega^2}{4(\cos\theta-\sin\theta)} \;.
\end{equation}
Since by definition ${\bf L}_{j}=I\ {\hat {\bf n}}_{j}\times{\partial_{t}{\hat {\bf n}}_{j}}$, the corresponding  term in the rotor Hamiltonian is $\mathbf{L}^2/2I = I  ({\hat {\bf n}}_{j}\times{\partial_{t}{\hat {\bf n}}_{j}})^2/2 = I \omega^2/2$, and for the moment of inertia we get 
\begin{equation}
 I =  \frac{1}{2 (\cos\theta-\sin\theta)} \;.
 \end{equation}

 Here we shall note that ${\bf L}_{j} = m$, the magnetization of the spin. As a consequence, the moment of inertia is equal to the magnetic susceptibility of the ferroquadrupolar states.
 
Let us also mention that both our $K$ and $I$ agree with the values given in Ref. [10] of our paper.

\subsection{Schmidt decomposition and density matrix of the SU(2) singlet in the SU(3) symmetric state}
%\numberwithin{equation}{section}
\renewcommand{\theequation}{S.\arabic{equation}}

At the ferro-SU(3) point, the ground state of the bilinear-biquadratic $S=1$ chain is the fully symmetrical SU(3) irreducible representation. For $L$ sites, it is the Young-tableaux with $L$ boxes arranged in a row, denoted by $(L,0)$, and its dimension is $(L+2) (L+1)/2$. Since the SU(2) is a subgroub of the SU(3), the states within a given SU(3) irreducible representation can further be classified according the the irreducible representation of the SU(2) subgroup. The fully symmetrical representation $(L,0)$ consists of $S=0,2,\dots,L$ spin states for $L$ even and $S=1,3,\dots,L$ for $L$ odd. This has important consequences for the ground state of the bilinear biquadratic model as we go away from the SU(3) symmetric point (but keeping the SU(2) symmetry): as we enter the ferromagnetic phase, the  $S=L$ high spin multiplet becomes the ground state, while  for $\theta\rightarrow \theta_F+0^+$ the $S=0$ spin singlet is the ground state. Our aim is to find the Schmidt decomposition of the latter state using the tools of group theory.

For example, on two sites there are two irreducible representations: the antisymmetrical $(0,1)$ Young tableaux (two boxes arranged vertically) of dimension 3, which is at the same time an SU(2) triplet, and the symmetrical $(2,0)$ Young tableaux (two boxes arranged horizontally) of dimension 6, which contains the  $|1\bar1\rangle +|\bar11\rangle-|00\rangle$  SU(2) singlet (using the $S^z$ basis), and the SU(2) quintuplet, which contains the $|11\rangle$ $S^z=2$ highest weight state.  

The SU(3) group has 2 diagonal and 6 off-diagonal generators. The generators of the SU(2) subgroup are the diagonal $S^z$ and the off-diagonal $S^+$ and $S^-$ operators, and the remaining five generators of the SU(3) group can be represented by the
\begin{equation}
\left(
\begin{array}{c}
Q_j^{+2} \\
Q_j^{+} \\
Q_j^{0} \\
Q_j^{-} \\ 
Q_j^{-2}
\end{array}
\right)
=
\left(
\begin{array}{c}
S^+_j S^+_j \\
 - S^+_j S^z_j - S^z_j S^+_j \\
 \frac{1}{\sqrt{6}} \left(4 S^z_j S^z_j 
 -  S^-_j S^+_j - S^+_j S^-_j \right) 
  \\
  S^-_j S^z_j +  S^z_j S^-_j \\ 
 S^-_j S^-_j
\end{array}
\right)
\end{equation}
quadrupolar (rank--2 tensor) operators. They satisfy the
\begin{eqnarray}
\left[S^z_j,Q^q_j \right] &=& q Q^q_j \;, \nonumber\\
\left[S^\pm_j,Q^q_j \right] &=& \sqrt{6 - q(q\pm1)} Q^{q\pm 1}_j 
\end{eqnarray}
commutation relations, and the Hermitian conjugates are
\begin{equation}
  (Q^{q})^\dagger_j = (-1)^q Q^{-q}_j \;.
  \label{eq:Qherm}
\end{equation}

The scalar product
\begin{equation}
{\bf Q}_j \cdot {\bf Q}_{j'} =  
\frac{1}{2} \sum_{q=-2}^{2} (-1)^q Q_{j}^{q} Q_{j'}^{-q} 
\end{equation}
is SU(2) invariant and the ${\bf Q}_j \cdot {\bf Q}_{j'} + {\bf S}_j \cdot {\bf S}_{j'}$ is SU(3) invariant. In fact, to describe
representations formed by all the spins we can define the invariant
\begin{equation}
 C_2 = \frac{1}{4}  \left( {\bf Q} \cdot {\bf Q} + {\bf S} \cdot {\bf S} 
\right)\;,
\label{eq_SU3casimir_SS+QQ}
\end{equation}
which commutes with all the 8 global generators (the ${\bf S}=\sum_j {\bf S}_j$ and the ${\bf Q}=\sum_j {\bf Q}_j$ summed over sites) of the SU(3) group. This is one of the two Casimir operators of the SU(3) group (the other one is cubic in the generators). The irreducible representation $(n_1,n_2)$ is an eigenstate of the Casimir operator, % 
\begin{equation}
 C_2 |(n_1,n_2)\rangle 
 =\frac{1}{3} 
 \left(3 n_1+ 3 n_2+ n_1^2 + n_2 n_1 + n_2^2\right)
 |(n_1,n_2)\rangle \;.
 \label{eq_SU3casimir_ev}
\end{equation}
The irreducible representations of the SU(2) group are characterized by the  total spin value $S$ as ${\bf S} \cdot {\bf S} |S\rangle = S(S\!+\!1) |S\rangle$. 

The representations of SU(3) are made up of several different irreducible representations of SU(2). The states
in a representation can be generated by applying the $S$'s and $Q$'s to a single state. While the $S^z$, $S^+$, and $S^-$ operators lead only to states within the same SU(2) irreducible representation, the $Q^q$ operators may generate states with different $S$'s (but still in the same SU(3) irreducible representation), where the initial and final total spin can differ by $0$ or $\pm2$.

\subsubsection{Matrix elements of the quadrupole operators}

Here we calculate the matrix elements of the quadrupole operators between the states of the $(n,0)$ fully symmetrical SU(3) irreducible representation. These states can be labeled as $|(n,0),S,m\rangle$ (where $m$ is the eigenvalue of the $S^z$ operator). For convenience, we consider the $n$ even case, where $S=0,2,\dots,n$. We will  omit the $(n,0)$ labels from the states when it is not confusing, so below we often write $|S,m\rangle$ instead of $|(n,0),S,m\rangle$.

%For the 3 SU(2) generators the matrix elements are the usual
%% 
%\begin{align}
% S^z |S,m\rangle &= m |S,m\rangle \nonumber\\
% S^\pm |S,m\rangle &= \sqrt{(S\mp m)(S \pm m+1)} |S,m\!\pm\! 1\rangle \;.
%\end{align}
%%

Since the quadrupole operators are rank-2 tensors with respect to the SU(2) operators, the Wigner-Eckart theorem applies to them:
\begin{equation}
\langle S'm'|Q^q | Sm \rangle = 
\langle S'|| {\bf Q} || S \rangle
 \langle S'm'|S2mq \rangle \;.
\end{equation}
Here $m'=m+q$, the $\langle S'm'|S2mq \rangle$ are the Clebsch-Gordan coefficients, and the $\langle S'|| {\bf Q} || S \rangle$ is the irreducible matrix element we are after.
Furthermore, from the Wigner-Eckart theorem we also learn that
\begin{equation}
\frac{\langle S\!+\!2||{\bf Q}|| S\rangle^2}{
\langle S||{\bf Q}|| S\!+\!2\rangle^2}
=
\frac{2 S +1 }{ 2S+5} \;,
\label{eq:Qirredmat}
\end{equation}

To calculate the irreducible matrix element, we first apply the (\ref{eq_SU3casimir_SS+QQ}) Casimir operator, as from Eq.~(\ref{eq_SU3casimir_ev}) we know that the $|(n,0),S,m\rangle$ is an eigenstate with eigenvalue $n(n+3)/3)$: 
\begin{equation}
 \frac{n(n+3)}{3} \left|S,m\right\rangle = \frac{1}{4} \left[S(S+1) + {\bf Q} \cdot {\bf Q} \right] \left|S,m\right\rangle \;,
\end{equation}
so that
\begin{equation}
\left\langle S,m\right| {\bf Q} \cdot {\bf Q} \left|S,m\right\rangle =
 \frac{4}{3}n(n+3) - S(S+1) \;.
\end{equation}
Inserting the identity $\sum_{S,m} \left|(n,0),S,m\right\rangle 
\left\langle (n,0),S,m\right|$ between the ${\bf Q}$ operators in the dot product, and using the orthogonality relation of the Clebsch-Gordan coefficients, 
we get
\begin{align}
\frac{8}{3} n (n\!+\!3) - 2 S(S\!+\!1)  
 =&  \langle S || {\bf Q} ||S\!+\!2 \rangle^2
+ \langle S || {\bf Q} ||S \rangle^2
\nonumber\\
&+ \langle S || {\bf Q} ||S\!-\!2 \rangle^2 \;,
\label{eq:set1}
\end{align}
where $S=0,2,\dots,n-2,n$. This is a set of $(n/2)+1$ linear equations (note $\langle S||{\bf Q}|| S\!+\!2\rangle=0$ for $S=n$ and $\langle S || {\bf Q} ||S\!-\!2 \rangle=0 $ for $S=0$).

We can use the commutation relations between the quadrupole operators to find other equations for the
irreducible matrix elements. 
For example, let us consider
$\left[ Q^{2-},Q^{+} \right] = 2 S^{-} $. Following the same procedure as above, we get
\begin{align}
0 = & (S\!+\!1) (S\!+\!2) (2 S\!+\!3) 
\left( 2S - \langle S||{\bf Q}||S\!-\!2\rangle^2 \right)
\nonumber\\
& + 3 (2 S\!-\!3)(S\!+\!2) \langle S||{\bf Q}||S\rangle^2
\nonumber\\
&
+ S \left(2 S^2\!+\!S\!+\!3\right) 
 \langle S||{\bf Q}||S\!+\!2\rangle^2 
\;,
\label{eq:set2}
\end{align} 
%\end{widetext}
%
and this leads to another set of $(n/2)+1$ linear equations.
 Eqs.~(\ref{eq:set1}) and (\ref{eq:set2}) together provide a set of linear equations from which we get the square of the irreducible matrix elements as
\begin{align}
\langle S||{\bf Q}|| S\rangle^2 &=  
  \frac{ 2 S   (2S \! +\! 2) ( 2 n \! +\! 3)^2}{6(2 S\! -\! 1) (2 S\! +\! 3)} \;, 
\nonumber\\
\langle S||{\bf Q}|| S\!+\!2\rangle^2 &= \frac{(2S\! +\! 2) (2S\! +\! 4) (n\! -\! S) (n \!+\! 3\! +\! S)}{(2 S\! +\! 3)(2 S\! + \!1)} \;,
\nonumber\\
\langle S||{\bf Q}||S\!-\!2 \rangle^2 &= \frac{(2S\! -\! 2)2S (n\! -\! S \!+\!2) (n \!+\! 1\! +\! S)}{(2 S\! - \! 1)(2 S\! + \!1)} \;.
\label{eq:Q-irreducible-matrix-elements} 
\end{align}

\subsubsection{Schmidt decomposition}

We divide the system into a right ${\mathcal R}$ and a left ${\mathcal L}$ subsystem, with equal number of sites $n$. We will concentrate below in the case when the total number of sites $L=2n$ is a multiple of 4, since in this case the SU(2) singlet is also present in the symmetrical $(n,0)$ SU(3) irreducible representations of the subsystems.
The Schmidt decomposition of the SU(2) singlet of the $(L,0)$ SU(3) irreducible representation is given as
\begin{align}   
 \left|(L,0),0,0\right\rangle = 
 \sum_{S} \sum_{m=-S}^{S} 
 \alpha(n,S,m) 
 &\left|(n,0),S,m\right\rangle_{\mathcal L} \nonumber\\
 &\otimes 
 \left|(n,0),S,-m\right\rangle_{\mathcal R} \;,
 \label{eq:Schmidt0}
\end{align}
where the sum $\sum_{S}$ is over the even $S=0,2,\dots,n$.
Since $\left|(L,0),0,0\right\rangle$ is a singlet, it is annihilated by the $S^+ =  S^+_{\mathcal L} + S^+_{\mathcal R}$ operator, and this leads to $\alpha(n,S,m)=- \alpha(n,S,m+1)$.  
We can then write 
\begin{equation}   
 \left|0,0\right\rangle = \sum_{S,m} (-1)^m \alpha_S \left|S,m\right\rangle_{\mathcal L} \otimes \left|S,-m\right\rangle_{\mathcal R}  \;.
\end{equation}

\begin{widetext}
Next, we use that the $|(L,0),0,0\rangle$ state is an eigenvector of the 
\begin{equation}
   2 \left[ C_2({\mathcal L}+{\mathcal R}) - C_2({\mathcal L}) - C_2({\mathcal R}) \right] =   {\bf Q}_{\mathcal L} \cdot {\bf Q}_{\mathcal R}
+ {\bf S}_{\mathcal L} \cdot {\bf S}_{\mathcal R}
\label{eq:casab}
\end{equation}
Casimir operator with an eigenvalue $4n^2/3$:
\begin{align}   
\left( 
  \frac{4}{3}n^2- {\bf Q}_{\mathcal L} \cdot {\bf Q}_{\mathcal R}
- {\bf S}_{\mathcal L} \cdot {\bf S}_{\mathcal R}
\right) 
%&\nonumber\\
\sum_{S,m} (-1)^m \alpha_S \left|S,m\right\rangle_{\mathcal L} & \otimes \left|S,-m\right\rangle_{\mathcal R}  = 0 \;.
\end{align}
This gives a large set of linearly dependent equations. Multiplying from the left by $\left\langle S',S'\right|_{\mathcal R} \left\langle S',-S'\right|_{\mathcal L}$, the nonvanishing terms are
\begin{align}
0=&\alpha_S \left( \frac{4}{3} n^2 
 - \sum_{m=-S}^{S}  
(-1)^m \left[ 
 \left\langle S,S\right| {\bf Q}_{\mathcal L} \left|S,m\right\rangle\cdot 
  \left\langle S,-S\right| {\bf Q}_{\mathcal R} \left|S,-m\right\rangle 
 + \left\langle S,S\right| {\bf S}_{\mathcal L} \left|S,m\right\rangle \cdot 
 \left\langle S,-S\right| {\bf S}_{\mathcal R}  \left|S,-m\right\rangle \right]
 \right)
\nonumber\\
&- \alpha_{S+2} 
\sum_{m=-S-2}^{S+2}  
   (-1)^m 
   \left\langle S,S\right| {\bf Q}_{\mathcal L} \left|S\!+\!2,m\right\rangle  \cdot 
   \left\langle S,-S\right|  {\bf Q}_{\mathcal R} \left|S\!+\!2,-m\right\rangle 
\nonumber\\&
-  \alpha_{S-2}  
\sum_{m=-S+2}^{S-2}  
(-1)^m 
 \left\langle S,S\right| {\bf Q}_{\mathcal L} \left|S\!-\!2,m\right\rangle 
  \cdot 
  \left\langle S,-S\right| {\bf Q}_{\mathcal R}  \left|S\!-\!2,-m\right\rangle 
  \;.
\end{align}
Using the hermicity properties [Eq.~(\ref{eq:Qherm})] for the ${\bf Q}_{\mathcal R}$ operators we can eliminate the $(-1)^m$ factors, and using the orthogonality relation  for the Clebsch-Gordon coefficients, we get  
\begin{align}
0=&\alpha_S \left( \frac{8}{3} n^2   +2 S(S+1)  
- \langle S || {\bf Q} || S \rangle^2 
  \right)
%\nonumber\\&
- \alpha_{S+2} 
 \langle S || {\bf Q} || S+2 \rangle^2 
%\nonumber\\&
-  \alpha_{S-2} 
 \langle S || {\bf Q} || S-2 \rangle^2 
 \;.
\end{align}
\end{widetext}
Replacing the irreducible matrix elements for the $Q$ given by Eq.~(\ref{eq:Q-irreducible-matrix-elements}), we arrive to the following set of equations:
\begin{eqnarray}
0 &=&  2 \alpha_{ n \!-\!2}  - (2 n \!+\!1)  \alpha_{ n } \;,\nonumber\\
&\vdots&
\nonumber\\
0 &=&  (n\!-\!S\!+\!2) (n\!+\!S\!+\!1) (S\!-\!1)  S (2S\!+\!3) \alpha_{S\!-\!2} \nonumber\\
&& - (2S\!+\!1) \left[ S (S\!+\!1) ( 2 n^2 \!-\! 2 n \!+\! 2 S^2 \!+\! 2 S \!-\!3) -2 n^2\right] \alpha_{S} \nonumber\\
&& + (n\!-\!S) (n\!+\!S\!+\!3) (S\!+\!1) (S\!+\!2)  (2S\!-\!1) \alpha_{S+2} \;,
\nonumber\\
&\vdots&
\nonumber\\
0 &=& n \alpha_{ 0 } -  (n\!+\!3)   \alpha_{ 2} \;.
\end{eqnarray}
These equations are solved by the  
\begin{equation}
\alpha_S = \frac{n-S+2}{n+S+1}\alpha_{S-2}
\end{equation}
recursion, which gives
\begin{equation}
\alpha_{S} = 2^{S}  \frac{(n+1)![(n+S)/2]!}{(n+S+1)![(n-S)/2]!} \;,
\end{equation}
assuming that $\alpha_{0} = 1$.

The eigenvalues of the reduced density matrix are proportional to the square of the normalized $\alpha$'s:
\begin{equation}
 \rho_{S} = \frac{|\alpha_{S}|^2}{\sum |\alpha_{S}|^2} \;,
\end{equation}
and it can be expressed in a closed form as
\begin{equation}
\rho_{S_{\mathcal L}=S} = (2n+1)\left(\frac{2^S n! [(n+S)/2]!}{(n + S + 1)![(n-S)/2]!}\right)^2\;.
\end{equation}
This is the Eq.~(7) in the paper.

\quad
\subsubsection{Asymptotic expansion in the $L\rightarrow \infty$ limit}

It is now straightforward to get the asymptotic behavior of the eigenvalues of the reduced density matrix (i.e. entanglement spectrum):
\begin{equation}
 - \ln \rho_{S} = \ln \frac{L}{4} +\frac{3}{L} + \frac{2}{L} S(S+1) + \cdots \;.\\
\end{equation}
The spacing between the levels in the spectrum grows linearly with $S$, and each level is $2S+1$ fold degenerate.

The von Neumann ($S_{E}$) and R\'enyi ($S_{R}$) entropy are
\begin{align}
  S_{E} &= -\sum_S (2S\!+\!1) \rho_S \log \rho_S \nonumber\\
  &= \ln \frac{L}{4} + 1 +\frac{5}{3}\frac{1}{L} + \cdots  
\end{align}
and 
\begin{align}
  S_{R} &= \frac{1}{1-\alpha} \sum_S (2S\!+\!1) \rho_S^{\alpha} 
  \nonumber\\
  &= \ln \frac{L}{4} +\frac{\ln \alpha}{\alpha-1} +\frac{7\alpha-2}{3\alpha}\frac{1}{L} + \cdots\;.
\end{align}
As expected, for $\alpha\rightarrow 1$ the  $S_{R} \rightarrow S_{E}$. 

Finally, the bipartite spin fluctuations  ${\mathcal F}\!=\!\left\langle \left(S^{z}_{\mathcal L}\right)^{\!2}\right\rangle$ are, using the SU(2) symmetry, 
\begin{align}
  {\mathcal F} = \frac{1}{3} \sum_S (2S+1) S(S+1) \rho_S 
    = \frac{L}{6} + \frac{1}{6} +\frac{1}{6L} + \cdots \;.
\end{align}
This behavior is shown by the green line in Fig 3(a).

\end{document}